\documentclass{PoS}
\usepackage{epsfig,epstopdf}

\title{Constraint on the internal structure of a neutron star from Vela pulsar glitches}

\ShortTitle{Vela pulsar glitch constraint}

\author{\speaker{Nicolas Chamel}\\ 
        Institut d'Astronomie et d'Astrophysique, CP-226, Universit\'e Libre de Bruxelles, 
        1050 Brussels, Belgium\\
        E-mail: \email{nchamel@ulb.ac.be}}

\abstract{Pulsars are spinning extremely rapidly with periods as short as about $1.4$ milliseconds 
and delays of a few milliseconds per year at most, thus providing the most accurate clocks 
in the Universe. Nevertheless, sudden spin ups have been detected in some pulsars like the
emblematic Vela pulsar. These abrupt changes in the pulsar's rotation period have long been
thought to be the manifestation of a neutron superfluid permeating the inner crust of neutron
stars. However, the neutron superfluid has been recently found to be so strongly coupled to
the crust that it does not carry enough angular momentum to explain the Vela data. We
explore the extent to which pulsar-timing observations can be reconciled with the standard glitch
theory considering the lack of knowledge of the dense-matter equation of state.}

\FullConference{The Modern Physics of Compact Stars 2015\\
		30 September 2015 - 3 October 2015\\
		Yerevan, Armenia}


\begin{document}

\section{Introduction}

Pulsars, the compact remnants of gravitational core-collapse supernova explosions, are spinning very rapidly 
with extremely stable periods $P$ ranging from about 1.4~milliseconds to a few seconds~\cite{beck09}. The variations $\dot{P}\equiv dP/dt$ 
of the rotation period of some pulsars over time $t$ do not exceed $10^{-21}$, as compared to $10^{-18}$ for the most 
accurate atomic clocks~\cite{hinkley2013}. Still, some pulsars have been found to suddenly spin up. The ensuing ``glitches'' 
in their rotational frequency $\Omega$, ranging from $\Delta\Omega\slash \Omega\sim 10^{-9}$ to $\sim 10^{-5}$, are 
generally followed by a long relaxation lasting from days to years, and sometimes accompanied by an abrupt change of 
the spin-down rate from $\vert\Delta \dot\Omega\slash\dot\Omega\vert\sim 10^{-6}$ up to $\sim 10^{-2}$. At the time of 
this writing, 472 glitches have been detected in 165 pulsars~\cite{jod12}. One of the most emblematic glitching pulsars 
is Vela (PSR B0833$-$45). 

The Vela pulsar was discovered in October 1968 by astronomers from Sydney University~\cite{large68}. Its very short 
period of about 89 milliseconds provided very strong evidence for the identification of pulsars as rotating neutron stars rather 
than rotating or vibrating white dwarfs. Moreover, the fact that this pulsar lies in a supernova remnant confirmed the scenario 
of gravitational core collapse of massive stars proposed much earlier by Baade and Zwicky~\cite{bz34}. Between 24 February and 
3 March 1969, the Vela pulsar was found to spin more rapidly than before~\cite{rad69,rei69}: the rotational frequency had increased 
by $\Delta \Omega/\Omega \simeq 2\times 10^{-6}$. The increase in the spin-down rate was even larger 
$\Delta \dot\Omega/\dot\Omega\simeq 7\times 10^{-3}$. Soon afterwards, Malvin Ruderman proposed that the glitch\footnote{According to George Greenstein, 
the term ``glitch'' was coined by Ruderman~\cite{green83}.} was the manifestation of 
crustquakes~\cite{rud69}. The idea was the following. As the star spins down, the presence of a solid crust prevents any readjustment 
of the stellar shape. Stresses build up to the point at which the solid crust will break down. The resulting quake entails a sudden 
decrease in the moment of inertia, and an increase in the spin frequency by conservation of angular momentum. As David Pines pointed out 
in 1999~\cite{pines99}, the Vela quake would be a cataclysmic event since this would correspond to an
Earthquake of magnitude 17 in which the entire surface is shifted by about 15 meters! On the other hand, crustquakes should be very rare: 
this model predicted that no such event should be observed again in Vela in a human lifetime~\cite{smo70,baym71}. The detection of the first 
glitch a few months after the discovery of the Vela pulsar thus appeared as a very unlikely coincidence, as remarked by Ruderman himself~\cite{rud69}: 
``The proposed model does not, however, account for the remarkable fact that such an uncommon event should just happen to 
occur during the short time, less than a year, during which this pulsar has been observed.'' Freeman Dyson speculated that neutron-star quakes could 
be more frequent if one assumes that the crustal stress arises from the accumulation of volcanic ashes at the stellar surface~\cite{dyson69}. 
In the fall of 1971, a second glitch occurred in Vela thus definitively ruling out Ruderman's crustquake theory. Other scenarios were 
proposed like corequakes~\cite{pines72}, planetary perturbation~\cite{michel70} or magnetospheric instabilities~\cite{scargle71}, but none of 
them were really convincing (see, e.g. Ref.~\cite{pines74}). For instance, if Vela glitches arose from corequakes they should have been 
seen in X-rays~\cite{rud76}. The eccentric orbital motion of planets around Vela predicted a strict regularity of glitches that were later found to be inconsistent with observations. Finally, the absence of radiative and pulse profil changes during glitches ruled out a magnetospheric origin. 

Since no such phenomena had ever been observed in other celestial bodies, glitches had to do with specific properties of neutron stars. 
Cameron and Greenstein proposed that the Vela glitch was due to the onset of fluid instabilities, the core rotating faster than the crust  
assuming ``viscous effects to be unimportant''~\cite{cam69}. After one year, the spin-down rate of Vela relaxed to the value it had before the 
glitch. This very long relaxation time was interpreted as a strong evidence for neutron-star superfluidity~\cite{baym69}. Neutron-star superfluidity had been 
actually predicted long before the discovery of pulsars by Migdal in 1959~\cite{mig59}, and had been first studied by Ginzburg and Kirzhnits in 
1964~\cite{gk64}. In 1972, Packard suggested that glitches 
are related to the metastability of this superfluid~\cite{pac72}. In 1975, Anderson and Itoh advanced the seminal idea that glitches are 
triggered by the sudden unpinning of superfluid vortices in neutron-star crust~\cite{and75}. A rotating superfluid is threaded by a regular array 
of quantized vortex lines, each carrying a quantum $\hbar$ of angular momentum. Such vortices have been observed in various kinds of superfluids in 
the laboratory and are thus expected to be present in the interior of rotating superfluid neutron stars. Quantized vortices tend to arrange themselves 
on a regular triangular array. The surface density of vortices in a rotating fermionic superfluid is given by $\displaystyle n_v=2m\Omega/(\pi\hbar)$, 
where $m$ is the mass of the fermions, $\Omega$ the angular frequency and $\hbar$ the Planck-Dirac constant. For a neutron superfluid, we obtain 
$\displaystyle n_v ({\rm km}^{-2}) \sim 10^{14} /P({\rm s})$. The Vela pulsar is thus expected to contain about $10^{17}$ vortices. 
The neutron superfluid in a neutron star is supposed to be weakly coupled to the rest of the star by mutual friction forces and to thus follow its 
spin-down via the motion of vortices away from the rotation axis unless vortices are pinned to the crust. In this case, the superfluid could rotate 
more rapidly than the crust. The lag between the superfluid and the crust induces a Magnus force acting on the vortices thereby producing a crustal stress. 
At some point, the vortices will suddenly unpin, the superfluid will spin down and, by the conservation of angular momentum, the crust will spin up leading 
to a glitch. This scenario found support from laboratory experiments with superfluid helium~\cite{tsa80}. Due to (non-dissipative) mutual neutron-proton 
entrainment effects, neutron superfluid vortices in the core of a neutron star carry a fractional magnetic quantum flux~\cite{sed80}. As a consequence, the 
core neutron superfluid is strongly coupled to the crust due to electrons scattering off the magnetic field of the vortex lines~\cite{alp84}. For this reason, 
only the neutron superfluid permeating the inner crust of a neutron star have been generally thought to be responsible for Vela glitches. Alpar, Pines and collaborators 
extended this vortex-mediated glitch scenario to explain the postglitch relaxation by the creeping of vortices~\cite{alp85,alp93}. Ruderman argued that if the pinning is 
strong enough the solid crust will crack before vortices get unpinned~\cite{rud91} (see also Chapter 14 of Ref.~\cite{beck09}). The subsequent motion of crustal plates 
could naturally explain the observed increase of the spin-down rate. Carter and collaborators later showed that the lack of centrifugal buoyancy and stratification are 
sources of crustal stress even in the absence of vortex pinning~\cite{car00,cc06}. Whether or not glitches are triggered by crust quakes remains uncertain (see, e.g. 
Refs.~\cite{lrr,hask15} and references therein for a review of more recent developments). 

Regardless of the actual glitch triggering mechanism, Vela pulsar timing data have 
been used to constrain the crustal moment of inertia of a neutron star~\cite{alp93,dat93,lnk99}. The fact that the inferred crustal moment of inertia is $\sim 2-3\%$,  
as expected for a neutron star with a typical mass $M\sim 1.4 M_\odot$ ($M_\odot$ being the mass of the Sun), provided additional support to the vortex-mediated glitch 
scenario. On the other hand, this scenario has been recently challenged~\cite{and12,cha13,hoo15,new15} by the realization that despite the absence of viscous drag the 
crust can still resist the flow of the neutron superfluid due to Bragg scattering~\cite{cha04,cha05,cch05,cha12}. 

In this paper, the extent to which the crustal superfluid can explain Vela pulsar glitches given the current lack of knowledge of nuclear physics is more closely examined considering a set of different unified dense-matter equations of state, treating consistently all regions of a 
neutron star, and taking into account nuclear-physics uncertainties. 

\section{Global model of superfluid neutron stars}

As discussed in the previous section, Vela pulsar glitches are generally thought to arise from sudden transfers of angular momentum between the 
neutron superfluid permeating the neutron-star inner crust and the rest of the star. An elegant and generic covariant variational formalism for describing such 
multicomponent systems was developed by Brandon Carter and his collaborators using exterior calculus. Originally formulated in the relativistic context~\cite{car89,car01}, 
it was later adapted to the Newtonian framework~\cite{cha04,CCI,CCII,CCIII}. A concise presentation of this formalism can be found, e.g., in Ref.~\cite{gou06} 
(see also Ref.~\cite{and07}). In this paper, we shall be concerned with a simplified global model of superfluid neutron stars as discussed in Ref.~\cite{cc06}. 

The total angular momentum $J$ of the star is the sum of 
the angular momentum $J_{\rm s}$ of the superfluid and of the angular momentum  $J_{\rm c}$ of the rest of the star (this includes the solid crust and 
the liquid core). Because of entrainment effects, the angular momentum of each component is a superposition of both the angular velocity $\Omega_{\rm s}$ 
and the observed angular frequency $\Omega$ of the rest of star
\begin{equation}\label{1}
J_{\rm s}=I_{\rm ss} \Omega_{\rm s} + I_{\rm sc} \Omega \, ,
\end{equation}
\begin{equation}\label{2}
J_{\rm c}=I_{\rm cs} \Omega_{\rm s} + I_{\rm cc} \Omega \,,
\end{equation}
where $I_{\rm ss}$, $I_{\rm sc}=I_{\rm cs}\leq 0$ and $I_{\rm cc}$ are partial moments of inertia. The total angular momentum can thus be written as 
\begin{equation}\label{3}
J=J_{\rm s}+J_{\rm c}=I_{\rm s} \Omega_{\rm s} + I_{\rm c} \Omega \, ,
\end{equation}
where $I_{\rm s}=I_{\rm ss}+I_{\rm sc}$ and $I_{\rm c}=I_{\rm cs}+I_{\rm cc}$ are the moments of inertia of the superfluid and of the rest 
of the star respectively. The evolution of the total angular momentum is governed by the external torque $\Gamma_{\rm ext}\leq 0$ arising 
from the pulsar electromagnetic radiation: 
\begin{equation}\label{4}
\dot{J}\equiv \frac{dJ}{dt} = \Gamma_{\rm ext} \, .
\end{equation}
The superfluid and the rest of the star are coupled by 
mutual friction forces induced by the dissipative motion (creep) of quantized vortex lines. 
Therefore, the superfluid is subject to an internal torque $\Gamma_{\rm int}\leq 0$ so that 
\begin{equation}\label{5}
\dot{J}_s = \Gamma_{\rm int}\leq 0 \, ,
\end{equation}
and consequently 
\begin{equation}\label{6}
\dot{J}_c =\Gamma_{\rm ext}- \Gamma_{\rm int} \, .
\end{equation}

Let us denote (discontinuous) variations of some quantity $Q$ during a glitch by $\Delta Q$ and (continuous) 
variations of this quantity during the interglitch period by $\delta Q$. Since no radiative and pulse profil changes have been detected during 
glitches, it is reasonable to assume that the total angular momentum $J$ is conserved:  
\begin{equation}
\label{7}
\Delta J_{\rm s}=-\Delta J_{\rm c}\, .
\end{equation}
Substituting Eqs.~(\ref{1}) and (\ref{7}), we thus obtain 
\begin{equation}
\label{8}
I_{\rm s} \Delta \Omega_{\rm s}=-I_{\rm c} \Delta \Omega\, .
\end{equation}
During the interglitch period, the loss of superfluid angular momentum due to mutual friction forces embedded in Eq.~(\ref{5}) implies
\begin{equation}\label{9}
\delta J_s \leq 0 \, ,
\end{equation}
which can be equivalently expressed as  
\begin{equation}\label{10}
\delta\Omega_{\rm s} \leq -\frac{I_{\rm sc}}{I_{\rm ss}} \delta
\Omega \, ,
\end{equation}
using Eq.~(\ref{1}). 
Friction effects prevent a long term build up of too large a deviation of the superfluid angular velocity $\Omega_{\rm s}$ from the externally 
observable value $\Omega$. This means that the average over many glitches (denoted by $\langle ... \rangle$) of the change of relative angular velocity 
$\Omega_{\rm s}-\Omega$ should be approximately zero
\begin{equation}\label{11}
\langle\Delta\Omega_{\rm s}+\delta\Omega_{\rm s}\rangle \simeq \langle \Delta\Omega
+\delta \Omega\rangle \, .
\end{equation}
On the other hand, averaging over glitches occuring during a time $t$, we have 
\begin{equation}\label{12}
\langle \Delta\Omega \rangle = \sum_i \Delta \Omega_i\, , \hskip 0.5cm \langle \delta \Omega \rangle = t \dot \Omega\, .
\end{equation}
Combining Equations~(\ref{8}), (\ref{10}), (\ref{11}) and (\ref{12}) yields the following constraint~\cite{cc06}
\begin{equation}
\label{13}
\frac{(I_{\rm s})^2}{I I_{\rm ss}}\geq \mathcal{G} \, ,
\end{equation}
where $I=I_{\rm s}+I_{\rm c}$ is the total moment of inertia of the star. 
The coefficient $\mathcal{G}$ can be obtained from pulsar-timing data and is defined as
\begin{equation}
\mathcal{G}\equiv 2 \tau_c A_g \, ,
\end{equation}
where $\tau_c=\Omega/(2|\dot\Omega|)$ is the pulsar characteristic age 
and the glitch activity parameter $A_g$ is given by 
\begin{equation}
\label{15}
A_g=\frac{1}{t}\sum_i\frac{\Delta\Omega_i}{\Omega}\, .
\end{equation}

\section{Constraint on the crustal moment of inertia}

Equation~(\ref{13}) can be alternatively expressed as 
\begin{equation}
\label{14}
\frac{I_{\rm s}}{I}\geq \mathcal{G} \frac{\bar m_n^\star}{m_n} \, ,
\end{equation}
where $m_n$ is the neutron mass and $\bar m_n^\star$ is a suitably weighted mean effective neutron mass~\cite{cc06}, arising from Bragg scattering of 
unbound neutrons by neutron-proton clusters~\cite{cha04,cch05}, and defined as $\bar m_n^\star/m_n=I_{\rm ss}/I_{\rm s}$. Ignoring entrainment effects by setting $\bar m_n^\star=m_n$ and approximating $I_{\rm s}$ by the crustal moment of inertia 
$I_{\rm crust}$ leads to the following inequality~\cite{lnk99} 
\begin{equation}
\label{16}
\frac{I_{\rm crust}}{I}\geq \mathcal{G}  \, .
\end{equation}
This condition has been widely applied to put constraints on the global structure of neutron stars as well as on the equation of state of dense matter (see, e.g., 
Refs.~\cite{lat07,bao08,ozel13}). However, according to the calculations of Refs.~\cite{cha04,cha05,cha12}, 
the mean effective neutron mass $\bar m_n^\star \gg m_n$ is substantially larger than the bare neutron mass $m_n$  so that $I_{\rm ss}/I_{\rm s}\gg 1$ leading 
to a much more stringent constraint than that given by Eq.~(\ref{16}). 
The ratio $(I_s)^2/I_{ss} I$ appearing in Eq.~(\ref{13}) depends on the internal structure of the star. It can be conveniently decomposed in terms of the 
crustal moment of inertia as 
\begin{equation}
\label{17}
\frac{(I_{\rm s})^2}{I I_{\rm ss}}=\frac{I_{\rm crust}}{I_{\rm ss}}\left(\frac{I_{\rm s}}{I_{\rm crust}}\right)^2\frac{I_{\rm crust}}{I}\, .
\end{equation}
The fractional moments of inertia $I_{\rm ss}/I_{\rm crust}$ and $I_{\rm s}/I_{\rm crust}$ depend essentially only on the properties of the neutron-star crust, and are approximately given by $I_{\rm ss}\simeq 4.58 I_{\rm crust}$ and $I_{\rm s}\simeq 0.893 I_{\rm crust}$~\cite{cha13} assuming that vortex pinning is effective in all regions of the inner crust (in reality, the inertia of the neutron superfluid could be lower). For the ratio $I_{\rm ss}/I_{\rm s}=\bar m_n^\star/m_n$, we thus obtain 
the value $5.13$. Adopting these values, the inequality~(\ref{13}) can be expressed as a constraint on the fractional moment of inertia of the crust 
\begin{equation}
\label{18}
\frac{I_{\rm crust}}{I}\geq 5.74 \mathcal{G} \, .
\end{equation}

\begin{figure}
\begin{center}
\includegraphics[scale=0.5]{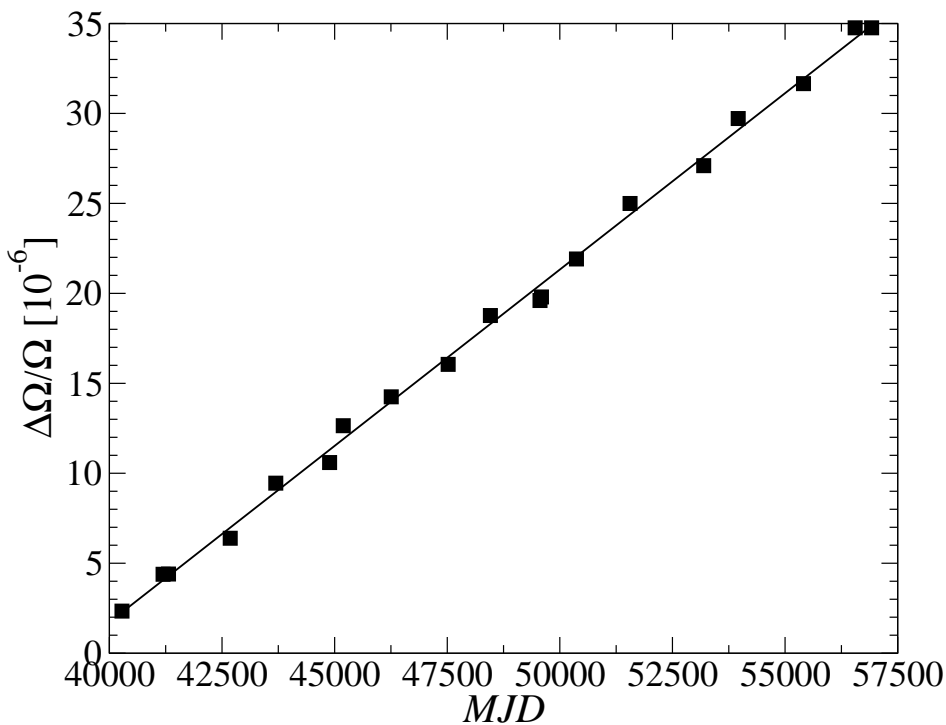}
\caption{Cumulated glitch amplitudes as a function of the modified Julian date for the Vela pulsar (PSR B0833$-$45) from Ref.~\cite{jod12} (square symbols) and 
linear fit (solid line).}
\label{vela}
\end{center}
\end{figure}

Because it was the first observed pulsar to exhibit glitches, Vela has become the testing ground for glitch theories. 
At the time of this writing, 19 glitches have been detected since 1969~\cite{jod12}. As shown in Fig.~\ref{vela}, the cumulated 
glitch amplitude increases almost linearly with time. A linear fit yields $A_g\simeq 2.27\times 10^{-14}$~s$^{-1}$. 
With the characteristic age $\tau_c=1.13\times 10^{4}$ years~\cite{atnf}, we find $\mathcal{G}\simeq 1.62\%$, and Eq.~(\ref{18}) thus becomes
\begin{equation}
\label{19}
\frac{I_{\rm crust}}{I}\biggr\vert_{\rm Vela}\geq 9.28 \% \, .
\end{equation}
Other glitching pulsars exhibit higher values for the coefficient $\mathcal{G}$. In particular, $\mathcal{G}\simeq 2.7\%$ for PSR~B1930$+$22 ~\cite{and12} leading to the 
more stringent constraint 
\begin{equation}
\frac{I_{\rm crust}}{I}\biggr\vert_{\rm PSR~B1930+22}\geq 16 \% \, .
\end{equation}
On the other hand, the statistical errors for this pulsar are larger than for Vela since only three glitches have been detected so far.

\section{Constraint on the mass and radius of neutron stars}

The fractional moment of inertia of the crust  $I_{\rm crust}/I$ depends on the global structure of the neutron star. We have computed this ratio 
using the following interpolating formula~\cite{lat00}: 
\begin{equation}
\label{20}
\frac{I_{\rm crust}}{I}=\frac{28 \pi P_{\rm cc} R^3}{3 M c^2}\frac{1-1.67 \beta-0.6 \beta^2}{\beta}\left[1+\frac{2 P_{\rm cc}}{\bar n_{\rm cc} m c^2}\frac{(1+7 \beta)(1-2 \beta)}{\beta^2}\right]^{-1}
\end{equation}
where $P_{\rm cc}$ and $\bar n_{\rm cc}$ are the pressure and average baryon number density at the crust-core interface respectively, $m$ is the baryon mass, 
$\beta=G M /(R c^2)$ ($G$ is the gravitational constant and $c$ the speed of light), $M$ and $R$ are respectively the gravitational mass and the circumferential
radius of a nonrotating neutron star. This formula was obtained by solving numerically Einstein's equations of general relativity for rotating neutron stars 
using a set of different dense-matter equations of state. 

The crust-core transition density and pressure were estimated~\cite{duc15} by the method described in Ref.~\cite{duc07}, which was shown to be extremely 
accurate~\cite{pea12}. For this purpose, we employed the recent family of accurately calibrated Brussels-Montreal nuclear energy density 
functionals from BSk22 to BSk26~\cite{gcp13}. These functionals are based on generalized Skyrme effective interactions, whose parameters were primarily fitted to 
all measured atomic masses with neutron number $N \geq 8$ and proton number $Z \geq 8$ from the 2012 Atomic Mass Evaluation~\cite{ame12}. 
Theoretical masses were calculated using the self-consistent Hartree-Fock-Bogoliubov method (see, e.g. Ref.~\cite{cpfdgp15} for a short review). 
With a root-mean-square deviation of about $0.5$~MeV, the Brussels-Montreal mass models are the most accurate microscopic mass models. A set of 
different functionals was constructed by imposing different values for the symmetry energy at saturation, and by fitting different realistic 
neutron-matter equations of state spanning different degrees of stiffness corresponding to current predictions of various microscopic calculations. 
A number of additional constraints were imposed (see Ref.~\cite{gcp13}). Although these functionals were not directly fitted to realistic equations of state of symmetric nuclear matter, they were found to be compatible with the constraints inferred from the analysis of heavy-ion collision experiments~\cite{dan02,lynch09}. Besides, the predicted values for the symmetry energy and its slope at saturation are consistent with those inferred from various experimental and theoretical constraints~\cite{tsa12,lat14}. For these reasons, the Brussels-Montreal functionals are well-suited for determining the crust-core transition in neutron stars. Neutron-star interiors are assumed to be in full thermodynamical equilibrium at zero temperature. 

\begin{figure}
\begin{center}
\includegraphics[scale=0.5]{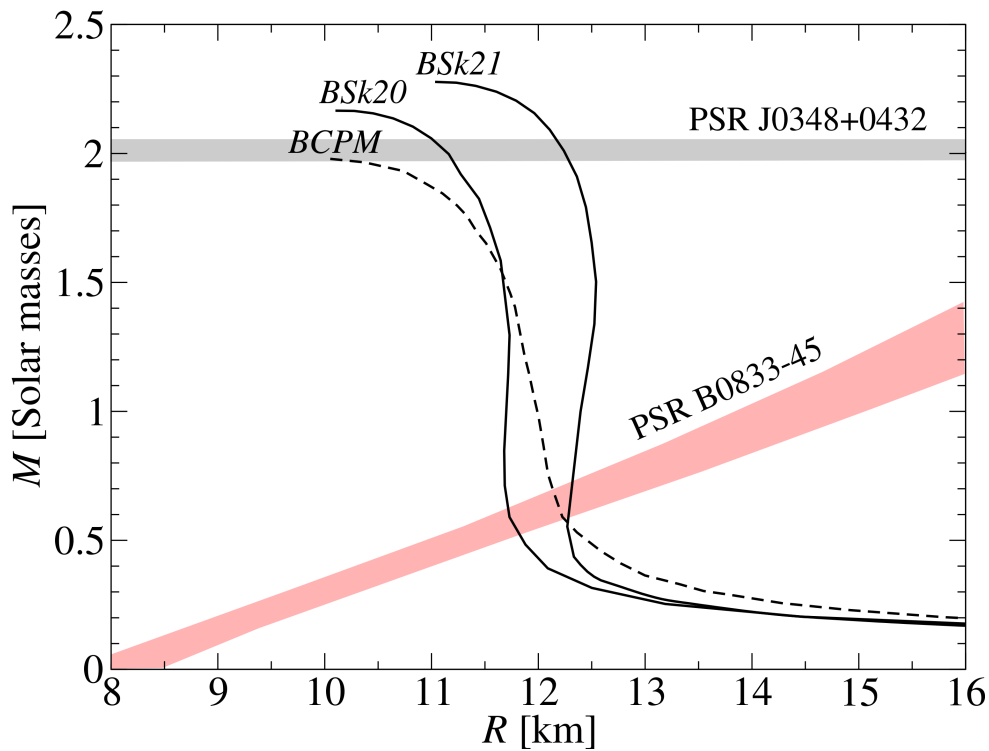}
\caption{Neutron-star mass-radius diagram for three different unified equations of state: the Brussels-Montreal equations of state BSk20 and BSk21~\cite{fant13,pot13}, and BCPM~\cite{bcpm}. The horizontal band represents the mass measurement of PSR~0348$+$0432~\cite{ant13}. The region lying above the red band labelled PSR~B0833$-$45 is excluded by pulsar-timing data if Vela glitches originate from the neutron superfluid in neutron-star crusts. See text for details.}
\label{constraint}
\end{center}
\end{figure}

Combining Eqs.~(\ref{19}) and (\ref{20}) for each set of crust-core transition density and pressure (as predicted by a given functional) leads to a specific mass-radius 
relation. These different relations are represented by the red band in Figure~\ref{constraint}. According to the standard vortex-mediated glitch theory, the 
region above this band is excluded by Vela pulsar glitch data. As an illustration, we have plotted the masses and radii of neutron stars, as predicted by three different unified equations of states, treating consistently all regions of a neutron star using the \emph{same} nuclear model from the surface to the core: BSk20 and BSk21~\cite{fant13,pot13}, and BCPM~\cite{bcpm}. We have 
also indicated the recent measurement of the mass of PSR~0348$+$0432~\cite{ant13}. 

As shown in Fig.~\ref{constraint}, the inferred mass of the Vela pulsar is at most $M\simeq 0.7 M_\odot$. On the other hand, the minimum mass of a neutron 
star formed from the gravitational core collapse of stars with a mass $M>8M_\odot$ is expected to lie in the range $0.88-1.28 M_\odot$~\cite{stro01}. Besides, 
the comparison of neutron-star cooling simulations with the observational estimates of the age and thermal luminosity of the Vela pulsar suggests that this neutron 
star is rather massive (see, e.g., Ref.~\cite{pot15}). This is the glitch puzzle~\cite{and12,cha13,hoo15,new15}. It has been recently argued that this puzzle could be 
resolved by invoking the lack of knowledge of the dense-matter equation of state~\cite{piek14,ste15}. On the other hand, the equations of state predicting thick enough 
neutron-star crusts are found to be at variance with existing constraints from laboratory experiments and astrophysical observations. For instance, the equation of state 
NL3max considered in Ref.~\cite{piek14} is inconsistent with the analysis of matter flow and kaon production in heavy-ion collision experiments~\cite{dan02,lynch09}. As for the parametrization TFcmax considered in Ref.~\cite{piek14}, the predicted values for the symmetry energy and its slope at saturation lie well outside the range of values expected from various experimental and theoretical constraints~\cite{tsa12,lat14}. This presumably explains why this model yields thick enough crust since the crust-core boundary is correlated with the slope of the symmetry energy~\cite{duc11}.
As a matter of fact, the average baryon number density $\bar n_{\rm cent}$ at the center of the neutron stars lying 
in the red band shown in Fig.~\ref{constraint} is rather low: $\bar n_{\rm cent} \lesssim 0.32$~fm$^{-3}$ for BSk20, and $\bar n_{\rm cent}\lesssim 0.28$~fm$^{-3}$ for 
BSk21. 
At such densities, the neutron-star core is generally expected to contain nucleons and leptons only (see, e.g. Ref.~\cite{deb15} and references therein for a recent review about hyperons in neutron stars). Moreover, the equation of state can be probed by laboratory experiments up to several times the density $n_0\simeq 0.16$~fm$^{-3}$ of symmetry nuclear matter at saturation. For instance in Refs.~\cite{dan02,lynch09}, the equation of state was constrained by Au+Au collision experiments at densities up to $4.5 n_0$, well above the density $\bar n_{\rm cent}\lesssim 2 n_0$ at the center of low-mass neutron stars. On the whole, although the equation of state of  dense matter remains uncertain, the constraints imposed by both laboratory experiments and astrophysical observations lead to the  prediction that the Vela pulsar is a low mass neutron star assuming that glitches originate from the superfluid in the crust.

\section{Conclusions}

The giant frequency glitches observed in the Vela pulsar provide strong evidence for the presence of superfluids in the interior of neutron stars. Predicted before the discovery of pulsars, neutron-star superfluidity has recently found additional support from the rapid cooling of the neutron star in Cassiopeia A~\cite{pag11,sht11}, from observations of the initial cooling in persistent soft X-ray transients~\cite{shternin07,brown09}, and from measurements of pulsar braking indices~\cite{alp06,ho12} (for a recent review of neutron-star superfluidity, see, e.g. Ref.~\cite{page14}). 

According to the standard theory~\cite{alp85}, Vela pulsar glitches are triggered by the sudden unpinning of neutron superfluid vortices in the crust of the star. Assuming that this scenario is correct, a constraint on the crustal moment of inertia of Vela can be inferred from timing data~\cite{dat93,alp93,lnk99}. Although giant glitches have been observed in other pulsars, observations of the Vela pulsar yield the most stringent constraint: $I_{\rm crust}/I\geq 1.6\%$. This constraint has been recently revised after realizing that the crustal superfluid is strongly entrained by the rest of the star due to Bragg scattering~\cite{cha13}: $I_{\rm crust}/I\geq 9.3\%$. Considering current experimental and theoretical constraints on the dense-matter equation of state, we find that the crustal superfluid does not carry enough angular momentum to explain Vela pulsar glitches. Even if crustal entrainment is ignored, the standard vortex-mediated glitch scenario has been  challenged by the observation in 2010 of a 
huge glitch in PSR~2334$+$6 from which the following constraint was inferred: $I_{\rm crust}/I\geq 9.4\%$~\cite{alp11}. These observations suggest that the neutron superfluid in the core of a neutron star contributes to glitches~\cite{gug14}.

\acknowledgments
This work was mainly supported by Fonds de la Recherche Scientifique - FNRS (Belgium). This work was also partially supported by the 
Simons Foundation (USA) and the European Cooperation in Science and Technology (COST) Action NewCompStar MP1304. The hospitality of the Aspen Center for Physics 
(Colorado, USA), where some of this work was carried out, is gratefully acknowledged. The Aspen Center for Physics is supported by National Science Foundation grant 
PHY-1066293. The author thanks Ali Alpar, Dany Page and Dmitry Yakovlev for discussions.


\begin{thebibliography}{99}
\bibitem{beck09} W. Becker (Ed.), \emph{Neutron Stars and Pulsars}, \emph{Astrophysics and Space Science Library} {\bf 357}, Springer  2009.
\bibitem{hinkley2013} N. Hinkley, J. A. Sherman, N. B. Phillips, M. Schioppo, N. D. Lemke, K. Beloy, M. Pizzocaro, C. W. Oates, 
and A. D. Ludlow, \emph{An Atomic Clock with $10^{-18}$ Instability}, 
\emph{Science} {\bf 341} (2013) 1215.
\bibitem{jod12} C.~M. Espinoza, A.~G. Lyne, B.~W. Stappers, M. Kramer, 
\emph{A study of 315 glitches in the rotation of 102 pulsars}, 
\emph{Mon.Not.Roy.Astron.Soc.} {\bf 414} (2011) 1679 (2011) ; 
http://www.jb.man.ac.uk/pulsar/glitches.html
\bibitem{large68} M. I. Large, A. E. Vaughan and B. Y. Mills, 
\emph{A Pulsar Supernova Association?} 
\emph{Nature} {\bf 220} (1968) 340. 
\bibitem{bz34} W. Baade and F. Zwicky, 
\emph{Suyernovae and Cosmic Rays}, \emph{Phys.Rev. }{\bf 45} (1934) 138.
\bibitem{rad69} V. Radhakrishnan and R. N. Manchester, 
\emph{Detection of a Change of State in the Pulsar PSR~0833$-$45}, 
\emph{Nature} {\bf 222} (1969) 228.  
\bibitem{rei69} P. E. Reichley and G. S. Downs,
\emph{Observed Decrease in the Periods of Pulsar PSR~0833$-$45}, 
\emph{Nature} {\bf 222} (1969) 229. 
\bibitem{green83} G. Greenstein, \textit{Frozen Stars}, Freundlich books, New York (1983), p.93. 
\bibitem{rud69} M. Ruderman, \emph{Neutron Starquakes and Pulsar Periods}, \emph{Nature} {\bf 223} (1969) 597.
\bibitem{pines99} D. Pines, \emph{Pulsar Glitches: To What Extent Do These Probe Crustal Superfluidity, Core-Crust Coupling, 
and the Equation of State of Dense Neutron Matter?}, in \emph{Pulsar Timing, General Relativity and the Internal Structure 
of Neutron Stars}, Edited by Z. Arzoumanian, F. Van der Hooft, and E. P. J. van den Heuvel. Amsterdam, The Netherlands, 1999, p. 199.
\bibitem{smo70} R. Smoluchowski, \emph{Frequency of Pulsar Starquakes}, 
\emph{Phys. Rev. Lett.} {\bf 24} (1970) 923.
\bibitem{baym71} G. Baym and D. Pines, \emph{Neutron starquakes 
and pulsar speedup}, \emph{Ann. Phys.} {\bf 66} (1971) 816.
\bibitem{dyson69} F. Dyson, \emph{Volcano Theory of Pulsars}, 
\emph{Nature} {\bf 223} (1969) 486.
\bibitem{pines72} D. Pines, J. Shaham, M. Ruderman, \emph{Corequakes and the Vela Pulsar}, 
\emph{Nature Phys. Sci.} {\bf 237} (1972) 83. 
\bibitem{michel70} F. Curtis Michel, \emph{Pulsar Planetary Systems}, \emph{Astrophys. J. Lett.} {\bf 159} (1970) L25. 
\bibitem{scargle71} J. D. Scargle and F. Pacini, \emph{On the Mechanism of the Glitches in the Crab Nebula Pulsar}, 
\emph{Nature Physical Science} {\bf 232} (1971) 144.  
\bibitem{pines74} D. Pines, J. Shaham and M. Ruderman, \emph{Neutron Star Structure from Pulsar Observations}, in 
\emph{IAU proceedings} {\bf 53} (1974) 189.
\bibitem{rud76} M. Ruderman, 
\emph{Crust-breaking by neutron superfluids and the VELA pulsar glitches}, 
\emph{Astrophys. J.} {\bf 203} (1976) 213. 
\bibitem{cam69} A. G. W. Cameron and G. S. Greenstein, 
\emph{Spin Down Effects in Neutron Stars}, 
\emph{Nature} {\bf  222} (1969) 862.
\bibitem{baym69} G. Baym, C.~J. Pethick, and D. Pines, 
\emph{Superfluidity in Neutron Stars}, 
\emph{Nature} {\bf 224} (1969) 673.
\bibitem{mig59} A.~B. Migdal, 
\emph{Superfluidity and the moments of inertia of nuclei}, 
\emph{Nucl. Phys.} {\bf 13} (1959) 655.
\bibitem{gk64} V.~L. Ginzburg and D.~A. Kirzhnits, 
\emph{On the superfluidity of neutron stars}, 
\emph{Zh. Eksp. Teor. Fiz.}{ \bf 47} (1964) 2006.
\bibitem{pac72} R.~E. Packard, 
\emph{Pulsar Speedups Related to Metastability of the Superfluid Neutron-Star Core}, 
\emph{Phys. Rev. Lett.} {\bf 28} (1972) 1080.
\bibitem{and75} P.~W. Anderson and N. Itoh, 
\emph{Pulsar glitches and restlessness as a hard superfluidity phenomenon}, 
\emph{Nature} {\bf 256} (1975) 25.
\bibitem{tsa80} J.~S. Tsakadze and S.~J. Tsakadze, 
\emph{Properties of slowly rotating helium II and the superfluidity of pulsars}, 
\emph{J. Low Temp. Phys.} {\bf 39} (1980) 649.
\bibitem{sed80} D. M. Sedrakyan, K. M. Shakhabasyan, 
\emph{The magnetic fields of pulsars}, \emph{Astrofizika} {\bf 8} (1972) 557 ; 
\emph{On a mechanism of magnetic field generation in pulsars}, 
\emph{Astrofizika} {\bf 16} (1980) 727.
\bibitem{alp84} M.~A. Alpar, S.~A. Langer and J.A. Sauls, 
\emph{Rapid postglitch spin-up of the superfluid core in pulsars}, 
\emph{Astrophys. J.} {\bf 282} (1984) 533.
\bibitem{alp85} D. Pines and M.~A. Alpar, 
\emph{Superfluidity in neutron stars}, 
\emph{Nature} {\bf 316} (1985) 27 (1985).
\bibitem{alp93} M. A. Alpar, H. F. Chau, K. S. Cheng, D. Pines, 
\emph{Postglitch relaxation of the VELA pulsar after its first eight large glitches - A reevaluation with the vortex creep model}, 
\emph{Astrophys. J.} {\bf  409} (1993) 345.
\bibitem{rud91} M. Ruderman, 
\emph{Neutron Star Crustal Plate Tectonics. III. Cracking, Glitches, and Gamma-Ray Bursts}, 
\emph{Astrophys. J.} {\bf 382} (1991) 587. 
\bibitem{car00} B. Carter, D. Langlois, and D. M. Sedrakian, 
\emph{Centrifugal buoyancy as a mechanism for neutron star glitches}, 
\emph{Astron. Astrophys.} {\bf 361} (2000), 795.
\bibitem{cc06} N. Chamel and B. Carter, 
\emph{Effect of entrainment on stress and pulsar glitches in stratified neutron star crust}, 
\emph{Mon.Not.Roy.Astron.Soc.} {\bf 368} (2006) 796. 
\bibitem{lrr} N. Chamel and P. Haensel,
\emph{Physics of Neutron Star Crusts'}, 
\emph{Living Rev. Relativity} {\bf 11} (2008) 10. http://www.livingreviews.org/lrr-2008-10
\bibitem{hask15} B. Haskell \& A. Melatos, 
\emph{Models of pulsar glitches}, 
\emph{Int. J. Mod. Phys.} {\bf D 24} (2015) 530008. 
\bibitem{dat93} B. Datta and M. A. Alpar, 
\emph{Implications of the Crustal Moment of Inertia for Neutron Star Equations of State}, 
\emph{Astron. Astrophys.} {\bf 275} (1993) 210. 
\bibitem{lnk99} B. Link, R.~I. Epstein and J. M. Lattimer, 
\emph{Pulsar Constraints on Neutron Star Structure and Equation of State}, 
\emph{Phys. Rev. Lett.} {\bf 83} (1999) 3362.
\bibitem{and12} N. Andersson, K. Glampedakis, W.C.G. Ho, C.M. Espinoza, 
\emph{Pulsar Glitches: The Crust is not Enough}, 
\emph{Phys. Rev. Lett.} {\bf 109} (2012) 241103.
\bibitem{cha13} N. Chamel, 
\emph{Crustal Entrainment and Pulsar Glitches}, 
\emph{Phys. Rev. Lett.} {\bf 110} (2013) 011101.
\bibitem{hoo15} J. Hooker, W. G. Newton and Bao-An Li, 
\emph{Efficacy of crustal superfluid neutrons in pulsar glitch models}, 
\emph{Mon.Not.Roy.Astron.Soc.} {\bf 449} (2015) 3559. 
\bibitem{new15} W. G. Newton, S. Berger, and B. Haskell, 
\emph{Observational constraints on neutron star crust-core coupling during glitches}, 
\emph{Mon.Not.Roy.Astron.Soc.} {\bf 454} (2015) 4400. 
\bibitem{cha04} N. Chamel, \emph{Entrainment in Neutron-Star Crusts}, PhD thesis, Universit\'e Paris VI, France, 2004. 
\bibitem{cha05} N. Chamel, 
\emph{Band structure effects for dripped neutrons in neutron star crust}, 
\emph{Nucl.Phys.} {\bf A747} (2005) 109. 
\bibitem{cch05} B. Carter, N. Chamel, P. Haensel, 
\emph{Entrainment coefficient and effective mass for conduction neutrons in neutron star crust: simple microscopic models}, 
\emph{Nucl. Phys.} {\bf A748} (2005) 675. 
\bibitem{cha12} N. Chamel, 
\emph{Neutron conduction in the inner crust of a neutron star in the framework of the band theory of solids}, 
\emph{Phys. Rev. C} {\bf 85} (2012) 035801.
\bibitem{car89} B. Carter,
\emph{Covariant Theory of Conductivity in Ideal Fluid or Solid Media},
in \emph{Relativistic Fluid Dynamics} (C.I.M.E., Noto, May 1987)
ed.  A.M. Anile, \& Y. Choquet-Bruhat, \emph{Lecture Notes in Mathematics} {\bf 1385}
(Springer - Verlag, Heidelberg, 1989), pp. 1-64.
\bibitem{car01} B. Carter, 
\emph{Relativistic Superfluid Models for Rotating Neutron Stars},
in \emph{Physics of Neutron Star Interiors}, 
ed. D. Blaschke, N. K. Glendenning, A. Sedrakian, 
\emph{Lecture Notes in Physics} {\bf 578} (Springer, 2001) pp.54-96. 
\bibitem{CCI} B. Carter and N. Chamel, 
\emph{Covariant Analysis of Newtonian Multi-Fluid Models for Neutron Stars I: Milne-Cartan Structure and Variational Formulation}, 
\emph{Int. J. Mod. Phys. } {\bf D13} (2004) 291. 
\bibitem{CCII} B. Carter and N. Chamel, 
\emph{Covariant Analysis of Newtonian Multi-Fluid Models for Neutron Stars II: Stress-Energy Tensors and Virial Theorems}, 
\emph{Int. J. Mod. Phys. } {\bf D14} (2005) 717. 
\bibitem{CCIII} B. Carter and N. Chamel, 
\emph{Covariant Analysis of Newtonian Multi-Fluid Models for Neutron Stars II: Transvective, Viscous, and Superfluid Drag Dissipation}, 
\emph{Int. J. Mod. Phys. } {\bf D14} (2005) 749. 
\bibitem{gou06} E. Gourgoulhon, 
\emph{An introduction to relativistic hydrodynamics}, 
\emph{EAS Publications Series} {\bf 21} (2006) 43.
\bibitem{and07} N. Andersson and G. Comer, 
\emph{Relativistic Fluid Dynamics: Physics for Many Different Scales}, 
\emph{Living Rev. Relativity} {\bf 10} (2007) 1. http://www.livingreviews.org/lrr-2007-1
\bibitem{lat07} J. Lattimer and M. Prakash, 
\emph{Neutron star observations: Prognosis for equation of state constraints}, 
\emph{Phys. Rep. }{\bf 442} (2007) 109.
\bibitem{bao08} Bao-An Li, Li-Wen Chen, Che Ming Ko, 
\emph{Recent progress and new challenges in isospin physics with heavy-ion reactions}, 
\emph{Phys. Rep. }{\bf 464} (2008) 113.
\bibitem{ozel13} F. \"Ozel, 
\emph{Surface emission from neutron stars and implications for the physics of their interiors}, 
\emph{Rep. Prog. Phys.} {\bf 76} (2013) 016901.
\bibitem{atnf} R.~N. Manchester, G.~B. Hobbs, A. Teoh and M. Hobbs, 
\emph{The Australia Telescope National Facility Pulsar Catalogue}, 
\emph{Astron. J.} {\bf 129} (2005) 1993 ;  
http://www.atnf.csiro.au/research/pulsar/psrcat
\bibitem{lat00} J. Lattimer and M. Prakash, 
\emph{Nuclear matter and its role in supernovae, neutron stars and compact object binary mergers}, 
\emph{Phys.Rep.} {\bf 333} (2000) 121.
\bibitem{duc15} C. Ducoin, private communications. 
\bibitem{duc07} C. Ducoin, Ph. Chomaz, and F. Gulminelli, 
\emph{Isospin-dependent clusterization of neutron-star matter}, 
\emph{Nucl. Phys.} {\bf A789}, 403 (2007).
\bibitem{pea12} J. M. Pearson, N. Chamel, S. Goriely, and C. Ducoin, 
\emph{Inner crust of neutron stars with mass-fitted Skyrme functionals}, 
\emph{Phys. Rev. C} {\bf 85} (2012) 065803.
\bibitem{gcp13} S. Goriely, N. Chamel, J.~M. Pearson, 
\emph{Further explorations of Skyrme-Hartree-Fock-Bogoliubov mass formulas. XIII. The 2012 atomic mass evaluation and the symmetry coefficient}, 
\emph{Phys. Rev. C} {\bf 88} (2013) 024308. 
\bibitem{ame12} G. Audi, M. Wang, A.H. Wapstra, F.G. Kondev, M. MacCormick,
X. Xu, and B. Pfeiffer, 
\emph{The Ame2012 atomic mass evaluation}, 
\emph{Chin. Phys. C} {\bf 36} (2012) 1287.
\bibitem{cpfdgp15} N. Chamel, J.~M. Pearson, A.~F. Fantina, C. Ducoin, S. Goriely, A. Pastore, 
\emph{Brussels-Montreal Nuclear Energy Density Functionals, from Atomic Masses to Neutron Stars}, 
\emph{Acta Phys. Pol. B} {\bf 46} (2015) 349. 
\bibitem{dan02} P. Danielewicz, R. Lacey, W.~G. Lynch, \emph{Determination of the Equation of State of Dense Matter}, \emph{Science} {\bf 298}  (2002) 1592 
\bibitem{lynch09} W.G.Lynch, M.B. Tsang, Y. Zhang, P. Danielewicz, M. Famiano, ,Z. Li, A.W. Steiner, 
\emph{Probing the symmetry energy with heavy ions}, \emph{Progress in Particle and Nuclear Physics} {\bf 62} (2009) 427.
\bibitem{tsa12} M.~B. Tsang et al., 
\emph{Constraints on the symmetry energy and neutron skins from experiments and theory}, 
\emph{Phys. Rev. C }{\bf 86} (2012) 015803. 
\bibitem{lat14} J. Lattimer, 
\emph{Symmetry energy in nuclei and neutron stars}, 
\emph{Nucl. Phys.} {\bf A928} (2014) 276. 
\bibitem{fant13} A. F. Fantina, N. Chamel, J. M. Pearson, S. Goriely,  
\emph{Neutron star properties with unified equations of state of dense matter}, 
\emph{Astron. Astrophys.} \textbf{559} (2013) A128.
\bibitem{pot13} A.~Y. Potekhin, A.~F. Fantina, N. Chamel, J. M. Pearson, S. Goriely, 
\emph{Analytical representations of unified equations of state for neutron-star matter}, 
\emph{Astron. Astrophys.} \textbf{560} (2013) A48.
\bibitem{bcpm} B. K. Sharma, M. Centelles, X. Vi\~nas, M. Baldo, and G. F. Burgio, 
\emph{Unified equation of state for neutron stars on a microscopic basis}, 
\emph{Astron. Astrophys.} {\bf 584} (2015) A103. 
\bibitem{ant13} J. Antoniadis, P. C. C. Freire, N. Wex et al., 
\emph{A Massive Pulsar in a Compact Relativistic Binary}, 
\emph{Science} {\bf 340} (2013) 1233232. 
\bibitem{stro01} K. Strobel and M.~K. Weigel, 
\emph{On the minimum and maximum mass of neutron stars and the delayed collapse}, 
\emph{Astron. Astrophys.} {\bf 367} (2001) 582. 
\bibitem{pot15} A. Y. Potekhin, J. A. Pons, D. Page, 
\emph{Neutron Stars-Cooling and Transport}, 
\emph{Space Science Reviews} {\bf 191} (2015) 239. 
\bibitem{piek14} J. Piekarewicz, F.~J. Fattoyev, and C.~J. Horowitz, 
\emph{Pulsar glitches: The crust may be enough}, 
\emph{Phys. Rev. C} {\bf 90} (2014) 015803. 
\bibitem{ste15} A.W. Steiner, S. Gandolfi, F.~J. Fattoyev, and W.~G. Newton, 
\emph{Using neutron star observations to determine crust thicknesses, moments of inertia, and tidal deformabilities}, 
\emph{Phys. Rev. C} {\bf 91} (2015) 015804. 
\bibitem{duc11} C. Ducoin, J. Margueron, C. Provid\^encia, and I. Vida\~na, 
\emph{Core-crust transition in neutron stars: Predictivity of density developments}, 
\emph{Phys. Rev. C} {\bf 83} (2011) 045810. 
\bibitem{deb15} D. Chatterjee and I. Vidana, 
\emph{Do hyperons exist in the interior of neutron stars ?}, 
\emph{Eur. Phys. J. } {\bf A 52} (2016) 29.
\bibitem{pag11} D. Page, M. Prakash, J. M. Lattimer, A. M. Steiner, 
\emph{Rapid Cooling of the Neutron Star in Cassiopeia A Triggered by Neutron Superfluidity in Dense Matter}, 
\emph{Phys. Rev. Lett.} {\bf 106} (2011) 081101. 
\bibitem{sht11} P. S Shternin, D. G. Yakovlev, C. O. Heinke, W. C. G. Ho, D. J. Patnaude, 
\emph{Cooling neutron star in the Cassiopeia A supernova remnant: evidence for superfluidity in the core}
\emph{Mon.Not.Roy.Astron.Soc.} {\bf 412} (2011) L108. 
\bibitem{shternin07} P.~S. Shternin, D.~G. Yakovlev, P. Haensel \&  A.~Y. Potekhin, 
\emph{Neutron star cooling after deep crustal heating in the X-ray transient KS 1731-260}, 
\emph{Mon.Not.Roy.Astron.Soc.} {\bf 382} (2007) L43 (2007). 
\bibitem{brown09} E.~F. Brown and A. Cumming, 
\emph{Mapping Crustal Heating with the Cooling Light Curves of Quasi-Persistent Transients}, 
\emph{Astrophys. J.} {\bf 698} (2009) 1020.
\bibitem{alp06} M. A. Alpar and A. Baykal, 
\emph{Pulsar braking indices, glitches and energy dissipation in neutron stars}, 
\emph{Mon.Not.Roy.Astron.Soc.} {\bf 372} (2006) 489. 
\bibitem{ho12} W. C. G. Ho and N. Andersson, 
\emph{Rotational evolution of young pulsars due to superfluid decoupling}, 
\emph{Nature Phys.}{\bf 8} (2012) 787.
\bibitem{page14} D. Page, J. M. Lattimer, M. Prakash, and A. W. Steiner, \emph{Stellar Superfluids} in 
\emph{Novel Superfluids, volume 2}, Edited by Karl-Heinz Bennemann and John B. Ketterson, 
Oxford University Press, 2014, p.505. 
\bibitem{alp11} M. A. Alpar, \emph{The Largest Pulsar Glitch and the Universality of Glitch Behavior}, 
\emph{AIP Conf. Proc.} {\bf 1379} (2011) 166. 
\bibitem{gug14} E. G\"ugercino\u{g}lu and M. A. Alpar, 
\emph{Vortex Creep Against Toroidal Flux Lines, Crustal Entrainment, and Pulsar Glitches}, 
\emph{Astrophys. J. Lett.} {\bf 788} (2014) L11.
\end{thebibliography}
\end{document}